\documentclass[aps,prb,twocolumn,amsmath,amssymb,superscriptaddress]{revtex4-1}

\usepackage{graphicx}
\usepackage{color}
\usepackage{makecell}
\usepackage{xspace}
\usepackage{amsthm}
\usepackage{amsfonts}
\usepackage{amssymb,bbm}
\usepackage{longtable}
\usepackage{array}
\usepackage{booktabs}
\usepackage{dsfont}
\usepackage[braket, qm]{qcircuit}

\begin{document}
	
	\title{Quantum walks and Dirac cellular automata on a programmable trapped-ion quantum computer}

\author{C. Huerta Alderete}
\email[e-mail: ]{aldehuer@umd.edu}
\affiliation{Joint Quantum Institute, Department of Physics, University of Maryland, College Park, MD 20742, USA}
\affiliation{Instituto Nacional de Astrof\'isica, \'Optica y Electr\'onica, Calle Luis Enrique Erro No. 1, Sta. Ma. Tonantzintla, Pue. CP 72840, Mexico}
\author{Shivani Singh}
\affiliation{The Institute of Mathematical Sciences, C. I. T. Campus, Taramani, Chennai 600113, India}
\affiliation{Homi Bhabha National Institute, Training School Complex, Anushakti Nagar, Mumbai 400094, India}
\author{Nhung H. Nguyen}
\affiliation{Joint Quantum Institute, Department of Physics, University of Maryland, College Park, MD 20742, USA}
\author{Daiwei Zhu}
\affiliation{Joint Quantum Institute, Department of Physics, University of Maryland, College Park, MD 20742, USA}
\author{Radhakrishnan Balu}
\affiliation{U.S. Army Research Laboratory, Computational and Information Sciences Directorate, Adelphi, Maryland 20783, USA}
\affiliation{Department of Mathematics \& Norbert Wiener Center for Harmonic Analysis and Applications, University of Maryland, College Park, MD20742}
\author{Christopher Monroe}
\affiliation{Joint Quantum Institute, Department of Physics, University of Maryland, College Park, MD 20742, USA}
\author{C. M. Chandrashekar}
\affiliation{The Institute of Mathematical Sciences, C. I. T. Campus, Taramani, Chennai 600113, India}
\affiliation{Homi Bhabha National Institute, Training School Complex, Anushakti Nagar, Mumbai 400094, India}
\author{Norbert M. Linke}
\affiliation{Joint Quantum Institute, Department of Physics, University of Maryland, College Park, MD 20742, USA}

\date{\today}

\begin{abstract}
	The quantum walk formalism is a widely used and highly successful framework for modeling quantum systems, such as simulations of the Dirac equation, different dynamics in both the low and high energy regime, and for developing a wide range of quantum algorithms. Here we present the circuit-based implementation of a discrete-time quantum walk in position space on a five-qubit trapped-ion quantum processor. We encode the space of walker positions in particular multi-qubit states and program the system to operate with different quantum walk parameters, experimentally realizing a Dirac cellular automaton with tunable mass parameter. The quantum walk circuits and position state mapping scale favorably to a larger model and physical systems, allowing the implementation of any algorithm based on  discrete-time quantum walks algorithm and  the dynamics associated with the discretized version of the Dirac equation.
\end{abstract}

	
	\maketitle
\section{\label{sec1}Introduction}
Quantum walks (QWs) are the quantum analog of classical random walks, in which the walker steps forwards or backwards along a line based on a coin flip. In a QW, the walker proceeds in a quantum superposition of paths, and the resulting interference forms the basis of a wide variety of quantum algorithms, such as quantum search \cite{Childs2003p59,Ambainis2003p507,Shenvi2003p052307, Ambainis2007p210,Magniez2007p413}, graph isomorphism problems \cite{Douglas2008p075303,Gamble2010p052313,Berry2011p4}, ranking nodes in a network \cite{Paparo2012p444,Paparo2013p2773,Loke2017p25,Chawla2019p}, and quantum simulations, which mimic different quantum systems at the low and high energy scale \cite{Arrighi2016p3467, DiMolfetta2014p157, DiMolfetta2013p042301, Chandrashekar2013p2829, Chandrashekar2010p062340, Strauch2006p054302, DiMolffeta2016p103038, Mallick2019p0150012, Chandrashekar2015p10005, Mallick2017p85}. In the discrete-time QW (DQW) \cite{Aharonov2001p50, Tregenna2003p83}, a quantum coin operation is introduced to prescribe the direction in which the particle moves in position space at each discrete step. In the continuous-time QW (CQW) \cite{Farhi1998p915,Gerhardt2003p290}, one can directly define the walk evolution on position space itself using continuous time evolution. We focus on DQWs and their implementation on gate-based quantum circuits in this work.

DQWs can be realized directly on lattice-based quantum systems where position space matches the discrete lattice sites. Such implementations have been reported with cold atoms \cite{Perets2008p170506,Karski2009p174} and photonic systems \cite{Schreiber2010p50502,Peruzzo2010p1500, Broome2010p153602,Tamura2019p}. In trapped ions, a DQW has been implemented by mapping position space to locations in phase space given by the degrees of freedom associated with the harmonic motion of the ion in the trap \cite{Schmitz2009p090504,Zahringer2010p100503, Gerritsma2010p68}. All these physical implementations have followed an analogue or quantum simulation approach. However, implementing QWs on a circuit based system is crucial to explore the algorithm applications based on QWs. The implementation of a DQW on a three-qubit NMR system \cite{Ryan2005p062317} and a CQW on a two-qubit photonic processor\cite{Qiang2016p11511} are the only two circuit-based implementations reported to date. To implement DQWs on circuit-based quantum processors, its necessary to map the position space to the available multi-qubit states. The range of the walk is set by the available qubit number and gate depth. The term Quantum Cellular Automaton (QCA) describes a unitary evolution of a particle on a discretized space \cite{Meyer1996p551, Mallick2016p25779,Perez2016p012328}, as occurs with QWs. In this context, the one-dimensional Dirac cellular automaton (DCA) has been derived from the symmetries of the QCA showing how the dynamics of the Dirac equation emerges \cite{Basio2015p244,Perez2016p012328, Meyer1996p551,Mallick2016p25779, Kumar2016p012116}. 

\begin{figure}[b!]
	\includegraphics[width=\linewidth]{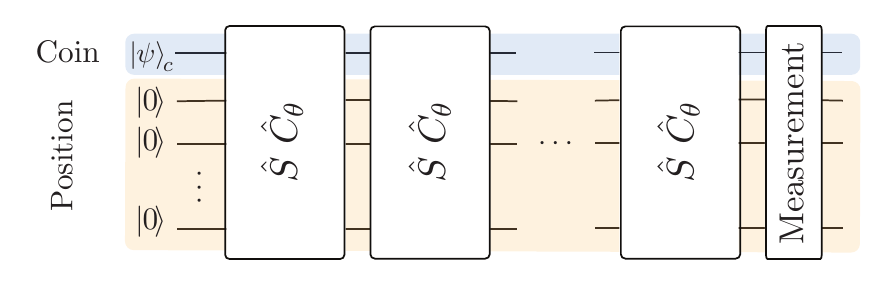}
	\caption{DQW scheme. Each step is composed of a quantum coin operation, $\hat{C}_{\theta}$, with tunable effective coin bias parameters, $\theta_{i}$, followed by a shift operation, $\hat{S}$.}
	\label{fig:QWscheme}
\end{figure}

Here we implement efficient quantum circuits for a DQW in one-dimensional position space, which provide the time-evolution up to five steps. We report the experimental realization of a DQW on five qubits within a seven-qubit programmable trapped-ion quantum computer \cite{Debnath2016p63}. With a tunable walk probability at each step we also show the experimental realization of a DCA where the coin bias parameter mimics the mass term in the Dirac equation. This will be central for discrete-time quantum simulation of the dynamics associated with the relativistic motion of a spin-1/2 particle in position space.

\section{\label{sec2} Review of quantum walks and the connection to the Dirac equation}

The DQW consists of two quantum mechanical systems, an effective coin and the position space of the walker, as well as an evolution operator, which is applied to both systems in discrete time-steps. The evolution is given by a unitary operator defined on a tensor product of two Hilbert spaces $\mathcal{H}_c \otimes  \mathcal{H}_p$ where,  $\mathcal{H}_c$ is the coin Hilbert space spanned by the internal states $\vert 0\rangle$ and $\vert 1\rangle$ of a single qubit, while $\mathcal{H}_p$ represents the position Hilbert space given by the position states $\vert x \rangle$ with $x  \in \mathbb{Z}$ encoded in several qubits as described below.
Here, the unitary quantum coin toss operation, $\hat{C}_{\theta}$, is a unitary rotation operator that acts on the coin qubit space, 
\begin{eqnarray}
\hat{C}_{\theta} = \left[ \begin{array}{cc}
\cos \theta &  -i\sin \theta \\
-i\sin \theta & \cos \theta
\end{array}\right]\otimes \hat{I}_{p},
\label{eq:qcoin}
\end{eqnarray}
where $\theta$ is a coin bias parameter that can be varied at each step to modify the QW path superposition weights. The conditional position-shift operator, $\hat{S}$, translates the particle to the left and right conditioned by the state of the coin qubit,
\begin{equation}\label{eq:shifto}
\hat{S}  = \vert 0 \rangle \langle 0 \vert \otimes \sum_{x\in\mathbb{Z}}  \vert x-1 \rangle \langle x \vert + \vert 1  \rangle \langle 1 \vert \otimes \sum_{x\in\mathbb{Z}} \vert x+1 \rangle \langle x \vert.
\end{equation}
The state of the particle in position space after $t$ steps of the walk, is accomplished by the repeated action of the operator $\hat{W}=\hat{S}\hat{C}_{\theta}$ on the initial state of the particle $\vert \psi \rangle_{c} = \alpha \vert 0 \rangle + \beta \vert 1 \rangle$ at position $x=0$, as shown in Fig. \ref{fig:QWscheme}, 
\begin{equation}\label{eq:evolution}
\vert \Psi(x,t)\rangle = \hat{W}^t \bigg[ \vert \psi \rangle_c \otimes \vert x=0\rangle  \bigg ] = \sum_x \left[\begin{array}{c}
\psi^{0}_{x, t} \\ \psi^{1}_{x, t}
\end{array}\right],
\end{equation}
where $\psi_{x,t}^{0(1)}$ denotes the left(right) propagating component of the particle at time-step $t$. The probability of finding the particle at position $x$ and time $t$ will be $P(x, t) = \vert \psi^{0}_{x, t}\vert^2 + \vert\psi^{1}_{x, t}\vert^2$.

Recent works have shown a relationship between DQWs and the Dirac equation \cite{Strauch2006p054302,Chandrashekar2013p2829,Arrighi2016p3467,Chandrashekar2010p062340,DiMolfetta2013p042301,Basio2015p244}. Starting form a discrete time evolution operator and then moving from position space to momentum space, Dirac kinematics can be recovered from the diagonal terms of the unitary evolution operator for small momenta in the small mass regime \cite{Strauch2006p054302,Chandrashekar2013p2829,Arrighi2016p3467}. In contrast with these proposals in the Fourier frame, we focus our implementation on the probability distribution of the DQW, which is analogous to the spreading of a relativistic particle. To realize a DCA and recover the Dirac equation, a split-step quantum walk, one form of the DQW, is used \cite{Mallick2016p25779}. Each step of a split-step quantum walk is a composition of two half step evolutions with different coin biases and position shift operators,
\begin{equation}
\label{eq:DA}
\hat{W}_{ss} = \hat{S}_{+}\hat{C}_{\theta_{2}} \hat{S}_{-} \hat{C}_{\theta_{1}},
\end{equation}
where the coin operation $\hat{C}_{\theta_j}$, with $j=1,2$, is given in Eq.\,(\ref{eq:qcoin}). The split-step position shift operators are,
\begin{eqnarray}
\hat{S}_{-} &=&  \ket{0}\bra{0} \otimes \sum_{x\in\mathbb{Z}}  \ket{x-1}\bra{x}+\ket{1}\bra{1} \otimes \sum_{x\in\mathbb{Z}}\ket{x}\bra{x} \nonumber \\
\hat{S}_{+} &=& \ket{0}\bra{0} \otimes \sum_{x\in\mathbb{Z}}  \ket{x}\bra{x}+\ket{1}\bra{1} \otimes \sum_{x\in\mathbb{Z}}\ket{x+1}\bra{x}.
\end{eqnarray}
Following \cite{Mallick2016p25779,Kumar2016p012116}, the particle state at time $t$ and position $x$ after the evolution operation $\hat{W}_{ss}$  is described by the differential equation,
\begin{eqnarray}\label{eq:temp_evolution}
\frac{\partial}{\partial t} \left[\begin{array}{c}
\psi^0_{x,t} \\ \psi^1_{x,t} 
\end{array}\right] &=& \cos \theta_{2} \left[\begin{array}{cc}
\cos \theta_1  & -i \sin \theta_1 \\
i \sin \theta_1  & -\cos \theta_1
\end{array}\right] \left[\begin{array}{c}
\frac{\partial \psi^0_{x, t}}{\partial x} \\
\frac{\partial \psi^1_{x, t}}{\partial x}
\end{array}\right] \nonumber \\
&+& \left[ \begin{array}{cc}
\cos(\theta_1+\theta_2)-1 & -i \sin(\theta_1+\theta_2) \\
-i \sin(\theta_1+\theta_2) & \cos(\theta_1+\theta_2)-1
\end{array}\right] \left[ \begin{array}{c}
\psi^0_{x,t} \\ \psi^1_{x,t} 
\end{array}\right]. \nonumber \\
\end{eqnarray}
By controlling the parameters $\theta_1$ and $\theta_2$, the split-step quantum walk turns into the one-dimensional Dirac equations for massless and massive spin-$1/2$ particles \cite{Mallick2016p25779,Kumar2016p012116}. For instance, the massless particle Dirac equation can be recovered for $\cos(\theta_1 + \theta_2) = 1$. Thereby, Eq. \eqref{eq:temp_evolution} becomes $i \hbar\left[\partial_{t} - \cos \theta_{2} \left(\cos \theta_{1} \sigma_z + \sin\theta_{1} \sigma_y\right)\partial_{x}\right] \Psi(x,t)=0$, which is identical to the Dirac equation of a massless particle in the relativistic limit \cite{Thaller2013p}. 
In contrast, considering $\theta_1=0$ and a very small value of $\theta_2$ corresponds to the Dirac equation for particles with small mass \cite{Thaller2013p,Gerritsma2010p68} in the form $i\hbar \left[\partial_{t} - (1-\theta^{2}_{2}/2)\sigma_z\partial_{x} + i \theta_{2}\sigma_x\right]\Psi(x,t) \approx 0$.

At the same time, by choosing $\theta_1=0$, the quantum walk operator $\hat{W}_{ss}$ given in Eq.\,(\ref{eq:DA}) takes the form of the unitary operator for a DCA \cite{Mallick2016p25779}, 
\begin{equation}
\hat{W}_{ss} =   \begin{bmatrix}  
\cos(\theta_2)S_- & -i \sin(\theta_2) \mathbbm{1} \\
-i \sin(\theta_2)\mathbbm{1}  & ~~\cos(\theta_2)S_+
\end{bmatrix} = U_{DCA}. 
\end{equation}
Within this framework, $\theta_2$ determines the mass of the Dirac particle. The split-step DQW described by the operator $\hat{W}_{ss}$ is equivalent to the two period DQW with alternate coin operations, $\theta_1$ and $\theta_2$, when the alternate points in position space with zero probability are ignored \cite{Zhang2017p052351}. Therefore, all the dynamics of a DCA can be recovered from the DQW evolution using $\hat{W}$ and alternating the two coin operations. See Appendix \ref{app:Comparison} for a comparison between DCA and the explicit solution of the Dirac equation. Typical features of the Dirac equation in relativistic quantum mechanics, such as the Zitterbewegung \cite{Mallick2016p25779} and the Klein paradox \cite{Basio2013p032301}, are also dynamical features of the DCA, as well as the spreading of the probability distribution and the entanglement of localized positive-energy states. We note that these effects have also been shown in direct analog simulations of the Dirac equation with trapped ions \cite{Gerritsma2010p68} and BECs \cite{LeBlanc2013p073011}.

\section{Experimental DQW implementation}

To realize the DQW on a system of qubits one must pick a mapping of the particle position to  the qubit space. As shown in \cite{Singh2019p}, there is no unique way to map position states to multi-qubit states, so each circuit decomposition depends on the configuration adopted. A direct mapping of each walker position to one qubit in the chain mimicking the arrangement of the qubit array is inefficient in terms of qubit number and gates required (the former grows linearly and the latter quadratically with the position space size modeled). In order to minimize resource use, we take advantage of a digital representation to map the position space into a multi-qubit state and re-order it in such a way that the state $\vert 0 \rangle ~(\vert 1 \rangle)$ of the last qubit corresponds to even (odd) position numbers. This allows us to minimize the changes needed in the qubit space configuration during each step of the walk (see Table \ref{tab:tab1}). To implement a quantum walk in one-dimensional position Hilbert space of size $2^n$, $(n+1)$ qubits are required. One qubit acts as the coin and the other $n$ qubits mimic the position Hilbert space with $2^n -1$ positions of a symmetric walk about $\vert x=0 \rangle$. The coin operation is achieved by single-qubit rotations on the coin-qubit while the shift operators are realized by using the coin as a control qubit to change the position state during the walk.

We realize the walk on a chain of seven individual $^{171}$Yb$^{+}$ ions confined in a Paul trap and laser-cooled close to their motional ground state \cite{Debnath2016p63,Landsman2019p67}. Five of these are used to encode qubits in their hyperfine-split $^{2}S_{1/2}$ ground level. Single qubit rotations, or R gates, and two-qubit entangling interactions, or XX gates are achieved by applying two counter-propagating optical Raman beams to the chain, one of which features individual addressing (see Appendix \ref{app:experiment} for experimental details). We can represent up to 15 positions, including the initial position $\vert x=0 \rangle$.
\begin{table}[h!]
	\centering
	\begin{tabular}{cccccccccccccccc}     
		$\vert$ \text{position} $\rangle$ &-7 & -6& -5& -4& -3& -2& -1& 0& 1& 2& 3& 4& 5& 6& 7 \\
		\rotatebox{90}{$\vert$ \text{qubit basis} $\rangle$} &\rotatebox{90}{$ \quad \vert ~1~0~0~1~\rangle$} & \rotatebox{90}{$\quad \vert ~1~0~1~0~\rangle$} & \rotatebox{90}{$\quad \vert ~1~0~1~1~\rangle$} & \rotatebox{90}{$\quad \vert ~0~1~0~0~\rangle$} & \rotatebox{90}{$\quad \vert ~0~1~0~1~\rangle$} & \rotatebox{90}{$\quad \vert ~0~1~1~0~\rangle$} & \rotatebox{90}{$\quad \vert ~0~1~1~1~\rangle$} & \rotatebox{90}{$\quad \vert ~0~0~0~0~\rangle$} & \rotatebox{90}{$\quad \vert ~0~0~0~1~\rangle$} & \rotatebox{90}{$\quad \vert ~0~0~1~0~\rangle$} & \rotatebox{90}{$\quad \vert ~0~0~1~1~\rangle$} & \rotatebox{90}{$\quad \vert ~1~1~0~0~\rangle$} & \rotatebox{90}{$\quad \vert ~1~1~0~1~\rangle$} & \rotatebox{90}{$\quad \vert ~1~1~1~0~\rangle$} & \rotatebox{90}{$\quad \vert ~1~1~1~1~\rangle$} 
	\end{tabular}
	\caption{Mapping of multi-qubit states to position states.}
	\label{tab:tab1}
\end{table}

\begin{figure*}[t!]
	\centering
	\includegraphics[scale=1]{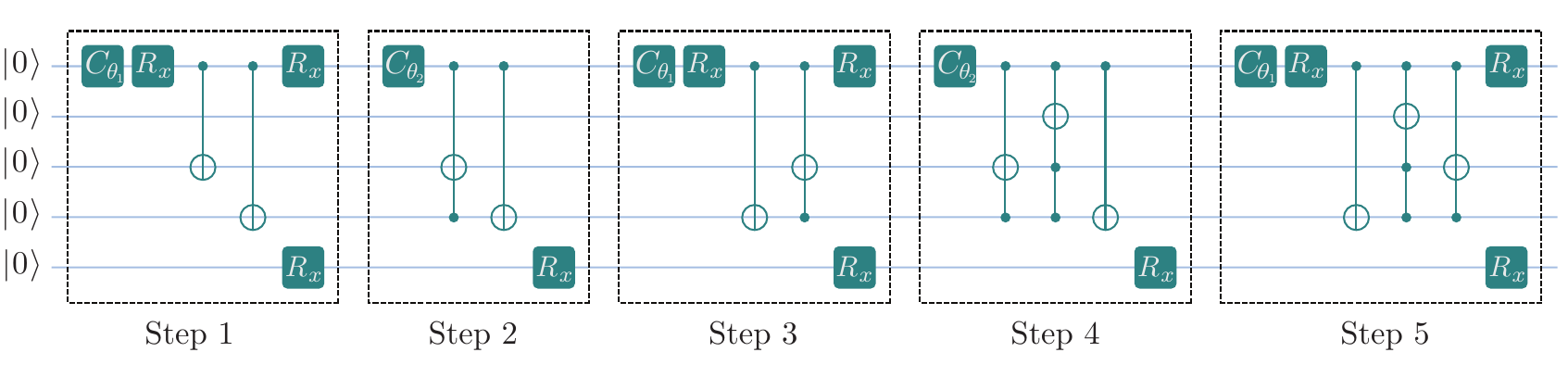}
	\caption{Circuit diagram for a DQW and DCA. Each dashed block describes one step in the quantum walk. }
	\label{fig:circuits}
\end{figure*}
Based on this position representation a circuit diagram for the DQW on five qubits with the initial state $\vert 0 \rangle \otimes \vert 0000 \rangle$ is composed for up to five steps, see Fig. \ref{fig:circuits}. Each evolution step, $\hat{W}$, starts with a rotation operation on the coin-qubit, $\hat{C}_{\theta_j}$, followed by a set of controlled gates that change the position state of the particle under $\hat{S}$. Due to the gratuitous choice of position representation used, it is enough to perform a single-qubit rotation on the last qubit at every step, which could also be done by classical tracking \cite{Singh2019p}.

Computational gates such as CNOT, Toffoli, and Toffoli-4 are generated by a compiler which breaks them down into constituent physical-level single- and two-qubit gates \cite{Debnath2016p63}. A circuit diagram detailing the compiled building blocks is shown in Appendix \ref{ap:gateblock}. To prepare an initial particle state different from $\vert 0 \rangle$ it is enough to perform a rotation on the coin-qubit before the first step. In some cases this rotation can be absorbed into the first gates in step one. Table \ref{tab:tab2} summarizes the number of native gates needed per step for initial state. To recover the evolution of the Dirac equation in a DQW after five steps, 81 single qubit gates and 32 XX-gates are required.
\begin{table}[!htbp]
	\centering
	\begin{tabular}{*9c ccc}
		\hline \hline
		 &  \multicolumn{7}{c}{DQW} && \multicolumn{3}{c}{DCA}\\
		\hline
		&&  \multicolumn{2}{c}{$\vert 0 \rangle / \vert 1 \rangle$} &&& \multicolumn{2}{c}{$\vert 0 \rangle+ i \vert 1 \rangle$} &&&  \multicolumn{2}{c}{$\vert 0 \rangle+ i \vert 1 \rangle$} \\
		\hline
		step && R & XX &&& R & XX &&& R & XX \\
		1 &&  5 &  2 &&& 6 &  2 &&& 5 &  2 \\
		2 && 10 &  4 &&& 10 &  4 &&& 12 &  4 \\
		3 && 12 &  4 &&& 12 &  4 &&& 11 &  4 \\
		4 && 25 & 11 &&& 25 & 11 &&& 27 & 11 \\
		5 && 26 & 11 &&& 26 & 11 &&& 26 & 11 \\
		\hline \hline
		Total: && 78 & 32 &&& 79 & 32 &&& 81 & 32
	\end{tabular}
\caption{Number of single- and two-qubit gates per step and total number of gates after a 5-step evolution.}
\label{tab:tab2}
\end{table}

\begin{figure}[t!]
	\centering
	\includegraphics[scale=1]{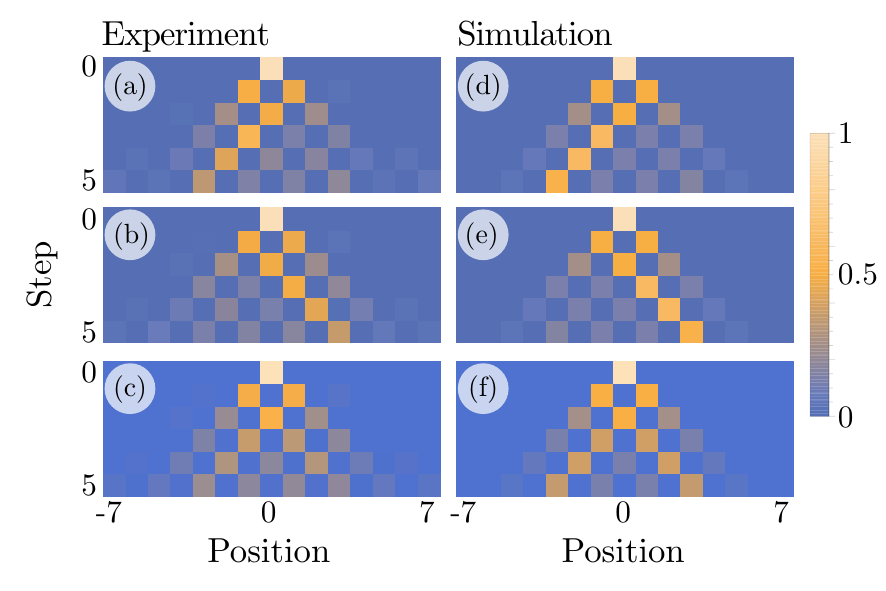}
	\caption{Discrete-time Quantum Walk. Comparison of the experimental results (left) and the theoretical quantum-walk probability distribution (right) for the first five steps with initial particle state (a)-(b) $\vert \psi\rangle_{c} = \vert 0\rangle$ , (c)-(d) $\vert \psi \rangle_{c} = \vert 1\rangle$, (e)-(f) $\vert \psi \rangle_{c} = \vert 0\rangle + i \vert 1\rangle$, and position state $\vert x = 0\rangle$.}
	\label{fig:DQWexp}
\end{figure}

After evolving a number of steps, we sample the corresponding probability distribution 3000 times and correct the results for readout errors.
For the DQW  evolution up to five steps shown in Fig. \ref{fig:DQWexp}, a balanced coin ($\theta_1 = \theta_2 = \pi/4$) is used where the initial position is $\vert x=0 \rangle $ for different initial particle states,  $\vert 0 \rangle$ in Fig. \ref{fig:DQWexp}(a), $\vert 1 \rangle$ in Fig. \ref{fig:DQWexp}(c), and an equal superposition of both in Fig. \ref{fig:DQWexp}(e). In Fig. \ref{fig:DQWexp}(b),(d), and (f) we show the ideal output from classical simulation of the circuit for comparison (see Appendix \ref{app:errors} for a plot of the difference).
With a balanced coin the particle evolves in equal superposition to the left and right position at each time step and upon measurement, there is a $50/50$ probability of finding the particle to the left or right of its previous position, just as in classical walk. If we let the DQW evolve for more than three steps before we perform a position measurement, we will find a very different probability distribution compared to the classical random walk \cite{Omar2006p042304}.

\begin{figure}[t!]
	\includegraphics[scale=1]{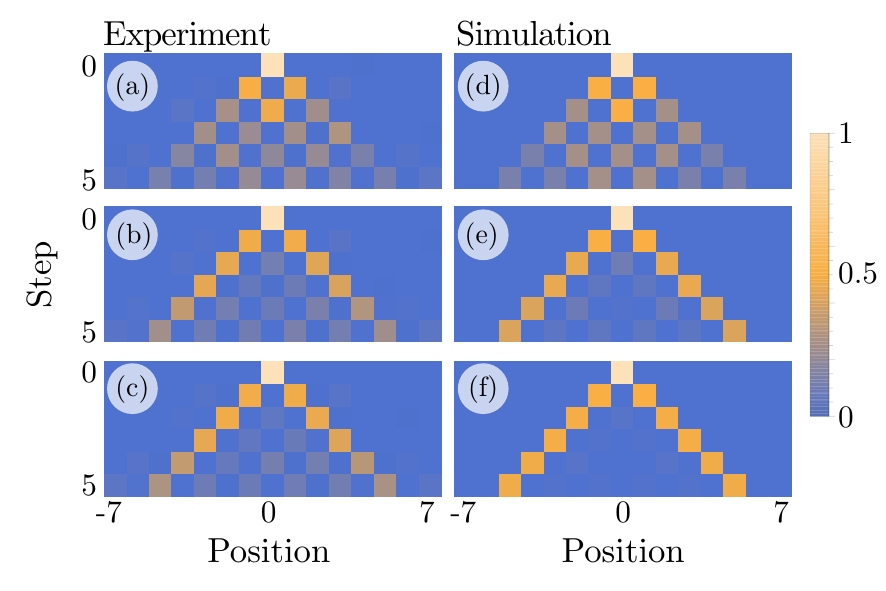}
	\caption{Output of a step-5 Dirac Cellular Automaton for $\theta_1 = 0$ and (a)-(b) $\theta_2 = \pi/4$, (c)-(d) $\theta_2 = \pi/10$ and (e)-(f) $\theta_2 = \pi/20$ with the initial state is $\vert\Psi_{in}\rangle = (\vert  0\rangle + i \vert 1\rangle) \otimes \vert x = 0\rangle$.}
	\label{fig:DCAexp}
\end{figure}

The same experimental setup can be used to recover a DCA with a two-period DQW. Here we set $\theta_1 = 0$ and varied $\theta_2$ to recover the Dirac equation for different mass values. In Fig. \ref{fig:DCAexp}, we show experimental results for $\theta_2 = \pi/4$, $\pi/10$ and $\pi/20$, corresponding to a mass 1.1357, 0.3305, and 0.159 in units of $\hbar/c^{2}s$, with the initial particle state in the superposition $|0\rangle + i|1\rangle$. The main signature of a DCA for small mass values is the presence of peaks moving outward and a flat distribution in the middle as shown for the cases with small values of $\theta_2$, Figs. \ref{fig:DCAexp}(b)-\ref{fig:DCAexp}(c). This bimodal probability distribution in position space is an indication of the one-dimensional analog of an initially localized Dirac particle, with positive energy, evolving in time which spreads in all directions in position space at speeds close to the speed of light \cite{Bracken2007p022322}. In contrast, a DCA with $\theta_2 = \pi/4$ corresponds to massive particle and hence there is a slow spread rather than a ballistic trajectory in position space.  

\section{\label{sec5} Concluding remarks}
We have shown how quantum walks form the basic elements for simulation of the dynamics associated with the free Dirac particle with positive energy. Despite the population mismatch of $0.05-0.2$ between the simulation and the experimental results after five steps, the final probability density exhibits the characteristic behavior of an initially localized Dirac particle. A key factor on the digitization of DQW/DCA is the mapping of qubit states to position space. An adequate mapping is important to minimize the number of gates on the protocol, and as a consequence, the resource scaling of the evolution. By increasing in the available number of qubits, these quantum circuits can be scaled to implement more steps and simulate a multi-particle DQW.
The number of gates has a polynomial growth rate with the number of steps \cite{FillionGourdeau2017p042343}.

\begin{acknowledgments}
	 The authors would like to thank to Y. Nam and C. Figgatt for helpful discussions. CHA acknowledges financial support from CONACYT doctoral grant no. 455378. CMC acknowledge the support from DST, Government of India under Ramanujan Fellowship grant no. SB/S2/RJN-192/2014 and US Army ITC-PAC contract no. FA520919PA139. NML acknowledges financial support from the NSF grant no. PHY-1430094 to the PFC@JQI. 
\end{acknowledgments}

\appendix

\section{Experimental details}\label{app:experiment}
The experiments are performed in a chain of seven individual $^{171}$Yb$^{+}$ ions confined in a Paul trap and laser-cooled close to their motional ground state \cite{Debnath2016p63,Landsman2019p67}. In order to guarantee higher uniformity in the ion spacing, matching the equally spaced individual addressing beams, the middle five of these are used to encode qubits in their hyperfine-split $^{2}S_{1/2}$ ground level, with an energy difference of $12.642821$ GHz. The two edge ions are neither manipulated nor measured, however, their contribution to the collective motion is included when creating the entangling operations.  The ions are initialized by an optical pumping scheme and are collectively read out using state-dependent fluorescence detection \cite{Olmschenk2007p052314}, with each ion being mapped to a distinct photomultiplier tube (PMT) channel. The system has two mechanisms for quantum control, which can be combined to implement any desired operation: single qubit rotations, or R gates, and two-qubit entangling interactions, or XX gates. These quantum operations are achieved by applying two counter-propagating optical Raman beams from a single 355-nm mode-locked laser \cite{Islam2014p3238}. The first Raman beam is a global beam applied to the entire chain, while the second is split into individual addressing beams, each of which can be controlled independently and targets one qubit. Single-qubit gates are generated by driving resonant Rabi rotations of defined phase, amplitude, and duration. Two-qubit gates are realized by illuminating two ions with beat-note frequencies near to the motional sidebands and creating an effective spin-spin (Ising) interaction via transient entanglement between the state of two ions and all modes of motion \cite{Choi2014p190502,Molmer1999p1835,Solano1999pR2539}. The average state detection fidelity for single- and two-qubit gate are 99.5(2)\% and 98-99\%, respectively. Rotations around the z-axis are achieved by phase advances on the classical control signals. Both the R as well as the XX angle can be varied continuously. State preparation and measurement (SPAM) errors are characterized and corrected by applying the inverse of an independently measured state-to-state error matrix \cite{Shen2012p053053}.

\section{Errors} \label{app:errors}

\begin{figure}[t!]
	\includegraphics[scale=1]{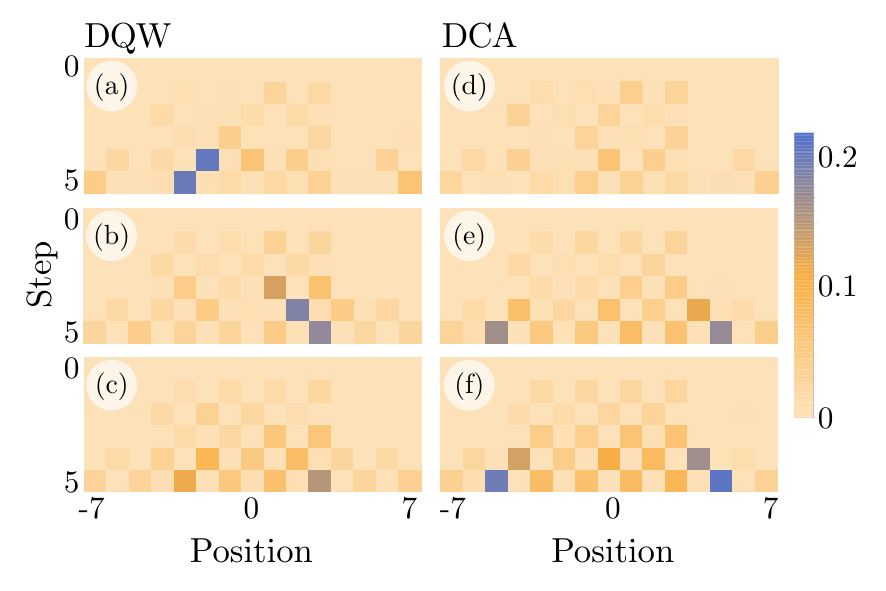}
	\caption{Experimental errors for DQW (left) and DCA (right).}
	\label{fig:error}
\end{figure}

In order to illustrate how our experiment performs, we plot the absolute value of the difference between the  measured and the simulated position distributions, Fig. \ref{fig:error}, they match the theoretical expectation closely. In both instances, DQW and DCA, the number of gates and hence the error incurred grows with the number of steps.
Apart from this, the output from the walk both, DQW and DCA, is designed to have zero probabilities for an alternate position, however, due to addressing crosstalk in the system, we see a small amount of population in these states.

\section{Comparison between Dirac kinematics and DCA} \label{app:Comparison}
\begin{figure}[h!]
	\includegraphics[scale=1]{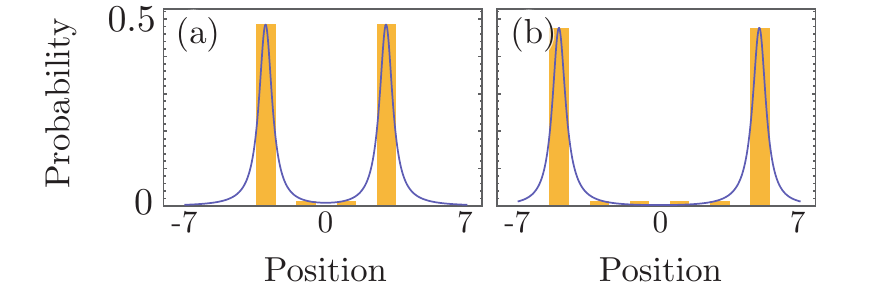}
	\caption{Numerical simulation of the explicit time-dependent solution of the one-dimensional Dirac equation (solid blue) and DCA (yellow bars) at (a) $t=3$ and (b) $t=5$ with $a=0.4$ and $\theta_{2}=\pi/20$.}
	\label{fig:Comparison}
\end{figure}

We use the explicit time-dependent solution of the one-dimensional Dirac equation provided by Strauch \cite{Strauch2006p054302}:
\begin{eqnarray}
	\Psi (x,t) = \frac{m \mathcal{N}}{ \pi} \left( \begin{array}{c}
		s^{-1} K_{1} (m s) \left[a + i (t+x)\right] + K_{0} (m s)\\
		s^{-1} K_{1} (m s) \left[a + i (t-x)\right] + K_{0} (m s)
	\end{array}\right),\nonumber\\
\end{eqnarray}
where $s = [x^{2} (a + i t)^{2}]^{1/2}$, $\mathcal{N}= \sqrt{(\pi / 2m)}[ K_{1} (2 m a)+  K_{0} (2 m a)]^{-1/2}$ the normalized factor and $K_{n}$ is the modified Bessel Function of order $n$, to show the corresponding probability density at time $t$ to the DCA after the time-step $t$, Fig. \ref{fig:Comparison}.
The relationship between the mass in the Dirac equation and the coin bias parameter is given by,
\begin{eqnarray}
	m \approx \frac{\theta_{2}}{1-\frac{\theta_{2}^{2}}{2}}.
\end{eqnarray}

\newpage
\onecolumngrid
\section{Gate block} \label{ap:gateblock}
The compiler breaks down the gate blocks shown in Fig. \ref{fig:circuits} (Toffoli-CNOT and Toffoli - Toffoli 4 - CNOT) into native R and XX gates as given by the following circuits, which are optimal in the XX-gate count.

\[
\Qcircuit @C=.7em @R=0.3em @!R {
	& \ctrl{2} & \ctrl{1} & \qw &&&&& \gate{R_y (\pi/2)} & \qw & \qw & \qw & \multigate{1}{XX(\pi/4)} & \gate{R_x(-\pi/2)} & \qw\\
	& \ctrl{1} & \targ & \qw & &=&&& \gate{R_y(\pi/2)} & \multigate{1}{XX(\pi/8)} & \gate{R_x(-\pi/4)} & \gate{R_y(-\pi/2)} & \ghost{XX(\pi/4)} & \gate{R_z(\pi/2)} & \qw\\
	& \targ & \qw & \qw &&&&& \qw & \ghost{XX(\pi/8)} & \qw & \qw & \qw  & \qw & \qw\\
}
\]
\[
\Qcircuit @C=.7em @R=0.3em @!R {
	&&&&&& &&&&& \qw & \qw & \qw  & \multigate{2}{XX(\pi/8)} & \gate{R_x(-\pi/4)} & \gate{R_y(-\pi/2)} & \qw \\
	&&&&&& &&\cdots &&& \multigate{1}{XX(-\pi/8)} & \gate{R_x(-\pi/4)} & \gate{R_y(-\pi/2)} & \ghost{XX(\pi/8)} & \qw & \qw & \qw \\
	&&&&&& &&&&& \ghost{XX(-\pi/8)} & \qw & \qw & \ghost{XX(\pi/8)} & \gate{R_x(-\pi/4)} & \qw & \qw\\
}
\]		

\[
\Qcircuit @C=.4em @R=0.3em @!R {
	& \ctrl{2} & \ctrl{3} & \ctrl{1} &\qw &&&&& \qw & \qw & \gate{R_z(5\pi/8)} & \gate{R_y(\pi/2)} & \multigate{1}{XX(\pi/4)} & \qw & \qw & \qw \\
	& \ctrl{1} & \ctrl{2} & \targ &\qw &&=&&& \gate{R_y(\pi/2)} & \multigate{2}{XX(-\pi/16)} & \gate{R_x(-3\pi/8)} & \gate{R_z(\pi/2)} & \ghost{XX(\pi/4)} & \gate{R_y(\pi/2)} &  \multigate{2}{XX(\pi/16)} & \qw \\
	& \targ & \ctrl{1} & \qw & \qw &&&&& \qw & \ghost{XX(-\pi/16)} & \qw & \qw & \qw & \qw & \ghost{XX(\pi/16)} & \qw  \\
	& \qw & \targ & \qw & \qw &&&&&  \qw & \ghost{XX(-\pi/16)} & \qw & \qw & \qw & \qw & \ghost{XX(\pi/16)} & \qw  \\
}
\]
\[
\Qcircuit @C=.4em @R=0.3em @!R {
	&&&&&&& &&&&& \qw & \qw & \multigate{1}{XX(\pi/4)} & \multigate{3}{XX(-\pi/16)} & \qw & \qw &  \qw & \qw & \\
	&&&&&&& &&\cdots&&& \gate{R_x(-5\pi/8)} & \gate{R_z(\pi/2)} & \ghost{XX(\pi/4)} & \ghost{XX(-\pi/16)} &  \gate{R_y(\pi/2)} & \gate{R_x(\pi/4)} & \qw & \qw\\
	&&&&&&& &&&&& \qw & \qw & \qw & \ghost{XX(-\pi/16)} & \gate{R_y(\pi/2)} & \multigate{1}{XX(\pi/8)} & \gate{R_x(\pi/4)} & \qw\\
	&&&&&&& &&&&& \qw & \qw & \qw & \ghost{XX(-\pi/16)} & \gate{R_x(\pi/8)} & \ghost{XX(\pi/8)} & \qw & \qw \\
}
\]
\[
\Qcircuit @C=.4em @R=0.3em @!R {
	&&&&&&& &&&&& \qw &\qw & \qw & \qw & \multigate{1}{XX(\pi/4)} & \qw & \qw &  \qw\\
	&&&&&&& &&\cdots&&& \qw & \qw & \multigate{1}{XX(\pi/8)} & \gate{R_z(\pi/2)} & \ghost{XX(\pi/4)} & \gate{R_y(\pi/2)} & \multigate{1}{XX(-\pi/8)} & \qw \\
	&&&&&&& &&&&& \gate{R_z(\pi/2)} & \gate{R_x(\pi/4)} & \ghost{XX(\pi/8)} & \qw & \qw & \qw & \ghost{XX(-\pi/8)} &  \qw\\
	&&&&&&& &&&&& \qw & \qw &  \qw & \qw & \qw & \qw & \qw & \qw \\
}
\]
\[
\Qcircuit @C=.4em @R=0.3em @!R {
	&&&&&&& &&&&&\qw & \qw & \multigate{2}{XX(\pi/8)} & \gate{R_y(-\pi/2)} & \qw & \qw & \qw & \qw  \qw  \\
	&&&&&&& &&\cdots&&& \gate{R_x(\pi/4)} & \gate{R_y(-\pi/2)} & \ghost{XX(\pi/8)} & \qw & \qw & \qw & \qw & \qw & \qw  \qw\\
	&&&&&&& &&&&&\qw & \qw & \ghost{XX(\pi/8)} & \gate{R_x(\pi/4)} &  \multigate{1}{XX(-\pi/8)} & \gate{R_x(\pi/4)} & \gate{R_y(-\pi/2)} & \qw \qw\\
	&&&&&&& &&&&& \qw & \qw & \qw & \qw & \ghost{XX(-\pi/8)} & \qw & \qw & \qw \qw \\
}
\]

\twocolumngrid

\end{document}